\documentclass[aps,pre,twocolumn,groupedaddress,showpacs,floatfix]{revtex4}
\usepackage{amsmath}
\usepackage{amssymb}
\usepackage{graphicx}

\begin{document}

\title{Reinforcing the Resilience of Complex Networks} 

\author{Luciano da Fontoura Costa} 
\affiliation{Institute of Physics of S\~ao Carlos. 
University of S\~ ao Paulo, S\~{a}o Carlos,
SP, PO Box 369, 13560-970, 
phone +55 162 73 9858,FAX +55 162 73
9879, Brazil, luciano@if.sc.usp.br}

\date{January 1, 2004}

\begin{abstract}   

Given a connected network, it can be augmented by applying a growing
strategy (e.g.  random or scale-free rules) over the previously
existing structure.  Another approach for augmentation, recently
introduced, involves incorporating a direct edge between any two nodes
which are found to be connected through at least one self-avoiding
path of length L.  This work investigates the resilience of random and
scale-free models augmented by using the three schemes identified
above.  Considering random and scale-free networks, their giant
cluster are identified and reinforced, then the resilience of the
resulting networks with respect to highest degree node attack is
quantified through simulations.  The results, which indicate that
substantial reinforcement of the resilience of complex networks can be
achieved by the expansions, also confirm the superior robustness of
the random expansion.

\end{abstract}


\maketitle

\section{Introduction}

A great deal has been leartn about several aspects of complex networks
\cite{Albert_Barab:2002, Newman:2003, Dorog_Mendes:2002} by looking at
such models from different theoretical and practical points of view,
such as network growth and critical phenomena
(e.g. \cite{Dorog_Mendes:2002, Erdos:1959}), node degree distribution
(e.g. \cite{Albert_Barab:2002}), distance between nodes
(e.g. \cite{Watts_Strogatz:1998}), diffusion
(e.g. \cite{Pastor:2001}), and resilience to attack, to name but a
few. As each of these situations drives the researcher to focus
attention on specific topological and functional aspects of the
investigated networks, they contribute to a more comprehensive and
integrated understanding of the many complexities of networks.  The
current work addresses the resilience issue by taking into account
three important issues frequently disregarded in the literature.
First, we target the situation where one wants to enhance an already
existing network with respect to attacks by adding a specific number
of edges; second, we consider the abrupt change of rules during the
network growth, producing \emph{hybrid} models; and third, we
investigate the potential of the recently introduced concept of
\emph{L-}expansion of a network \cite{Costa_expand:2004} for enhancing
resilience.

Because of its immediate practical consequences to Internet and
distributed systems, the problem of characterizing the resilience of
complex networks has received growing attention, especially after the
seminal papers by Albert et al. \cite{Albert_attack:2000}, who
addressed node deletion in scale-free models of Internet, and Callaway
et al.'s \cite{Callaway_etal:2000} investigation on random networks
under attack.  Other related works include Holme et al.'s
\cite{Holme_etal:2002} comprehensive comparative investigation of the
resilience of several types of networks considering different schemes
for attacking nodes and edges, and Cohen et al.'s analysis of internet
breakdown \cite{Cohen_etal:2000}.  Works targeting specific types of
network include, but are not limited to, Newman's investigation of
e-mail networks \cite{Newman:2002}, Jeong at al. study of metabolic
systems \cite{Jeong_etal:2000}, and Dunne's analysis of food webs
\cite{Dunne:2002}.  More recently, the concept of \emph{L-}expansions
of a complex network was suggested \cite{Costa_expand:2004} which, by
enhancing the network connectivity, was believed to present good
potential for increasing the resilience of existing networks.

This paper starts by reviewing the concept of \emph{L-}expansions and
augmentations of a network and follows by presenting the comparison
framework, the obtained results, and respective discussion.

\section{\emph{L-}Augmentations of a Network}

\begin{figure}
 \begin{center} 
   \includegraphics[scale=.45,angle=0]{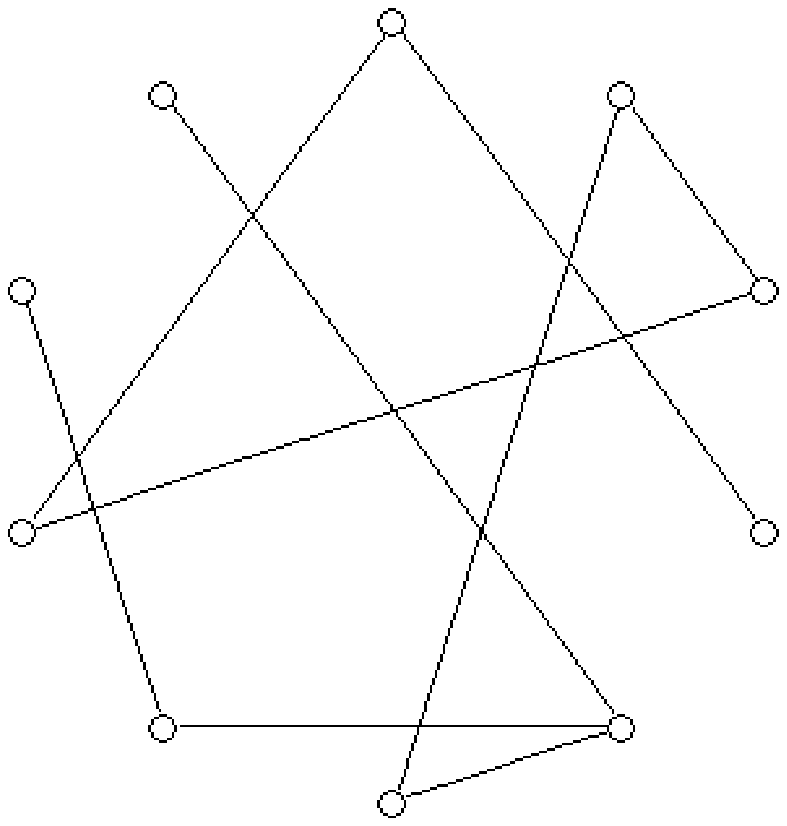} 
   \includegraphics[scale=.45,angle=0]{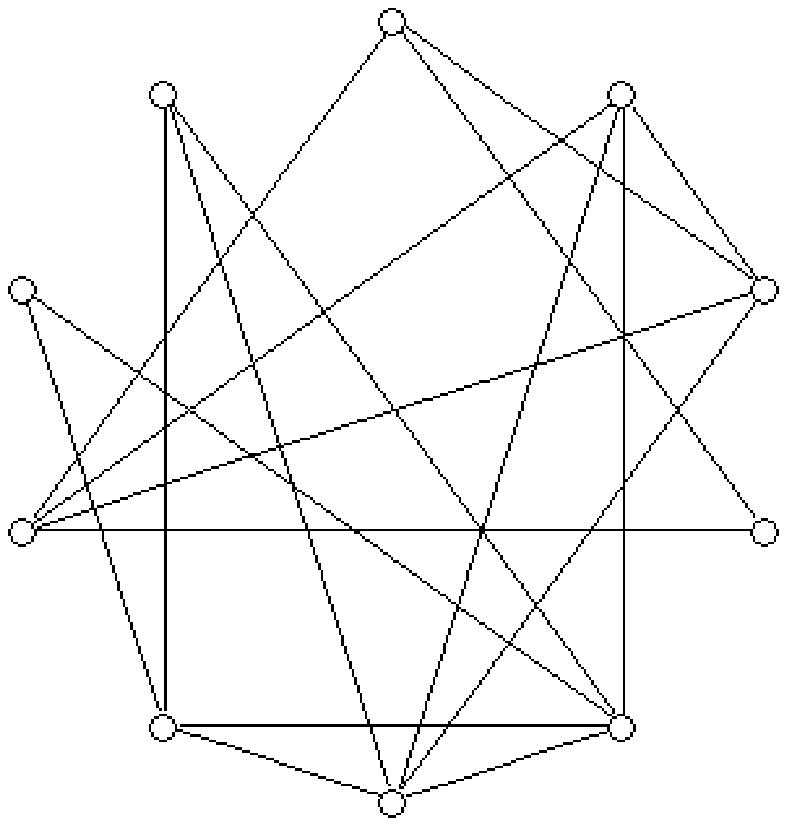} \\
   (b) \hspace{3cm}  (c) \\
   \includegraphics[scale=.45,angle=0]{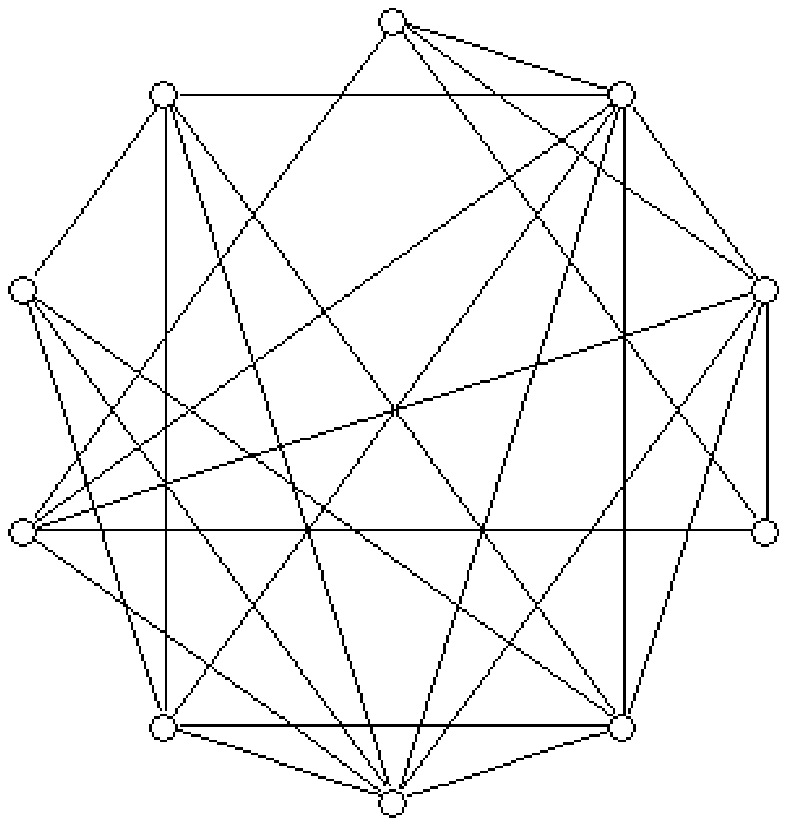} 
   \includegraphics[scale=.45,angle=0]{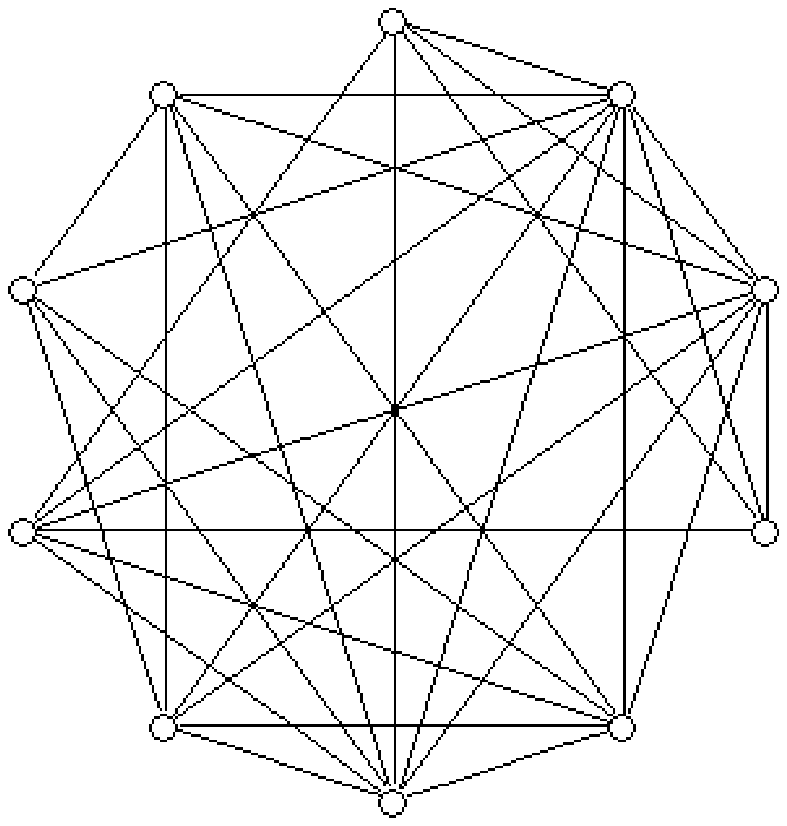} \\
   (d) \hspace{3cm}  (e) 

   \caption{A simple graph (a) and its respective 
   \emph{2-} (b), \emph{3-} (c) and 
       \emph{4-}augmentations (c).~\label{fig:ex}} 
 \end{center}
\end{figure}

Recently introduced \cite{Costa_expand:2004}, the concept of
\emph{L-}expansion of a given network (any type, directed or not)
seems to provide good potential for reinforcing the connectivity
regularity of existing networks, with implications for resilience.
Given a graph $\Gamma$, its \emph{L-}expansion consist of a graph
where connections from node $i$ to $j$ are established whenever there
exists a self-avoiding path (i.e. never passing by the same node
twice) of length $L$ connecting $i$ to $j$ in $\Gamma$.  Here we
introduce the concept of \emph{L-}augmented network in order to
express networks obtained by the union of the original graph with its
respective \emph{L-expanded} model. This simple concept is illustrated
in Figure~\ref{fig:ex}, which shows an original network (a) and its
respective \emph{2-}, \emph{3-} and \emph{4-}augmented versions.
Higher order expansions, which imply an exceedingly high number of
additional edges, are not considered in the current work.  It is
interesting to observe that these augmentations reinforce the
regularity of the network up to $L$ length.  An important global
measurement of the effect of the augmentation on the network
connectivity is the ratio between the number of connections in the
augmented and original networks, henceforth represented as $\rho$ and
denominated \emph{augmentation ratio}.

\section{Hybrid Networks}

The $N$ nodes of the network of interest $\Gamma$ are henceforth
represented as $k$ and the $E$ edges as ordered pairs $(i,j)$, with
the respective adjacency matrix being expressed as $A$.  No
self-connections are allowed.  Given a network $\Gamma_{\alpha}$ of a
specific type $\alpha$ (e.g. random or scale-free), its augmentation
(see, e.g., \cite{Frank:1994}) $\Pi_{\alpha}(\Gamma_{\beta})$ can be
obtained by applying the growing rules of any other model type $\beta$
\emph{over the existing network} $\Gamma$, in order to add an
additional number of edges $\Delta E$ without changing the number of
nodes.  Thus, we can have a random model augmented by the scale-free,
or a scale-free network followed by an \emph{L-}expansion.  Such
combinations of growing schemes are henceforth called \emph{hybrid
augmentation}, of which the current work considers the six following
situations:

\begin{trivlist}

  \item (i) random followed by random --- i.e. $\Pi_{R}(\Gamma_R)$; 
  \item (ii) random followed by scale-free --- i.e. $\Pi_{SF}(\Gamma_{R})$; 
  \item (iii) random followed by its \emph{L-}expansion --- i.e. 
     $\Pi_{LE}(\Gamma_{R})$; 
  \item (iv) scale-free followed by random --- i.e. $\Pi_{R}(\Gamma_{SF})$; 
  \item (v) scale-free followed by scale-free --- i.e. $\Pi_{SF}(\Gamma_{SF})$; 
  \item (vi) scale-free followed by its \emph{L-}expansion ---
   i.e. $\Pi_{LE}(\Gamma_{SF})$.  

\end{trivlist}

Observe that the two cases where a model is followed by an
augmentation of the same type are equivalent to considering a single
network of the same type containing the same number of nodes and edges
as in the other cases.  It is interesting to observe that the
augmentation of a network where $\alpha \neq \beta$ typically is
\emph{not} commutative, i.e. generally $\Pi_{\alpha}(\Gamma_{\beta})
\neq \Pi_\beta(\Gamma_{\alpha})$.  For the sake of a fair comparison
of the models, all networks derived from the initial connected graph
$\Gamma$ have the same number of nodes and edges.  More specifically,
the procedure for generating the hybrid augmented models starts by
growing a network of type $\alpha$ up to $P$ nodes, and the giant
cluster is identified, containing $No \leq P$ nodes and $Eo$
connections. The number of nodes is determined as $P= 2i/No/(No-1)$,
for $i=1,2,3$.  This connected network acts as the original network
$\Gamma$, which is subsequently augmented by $\Delta E$ new edges
according to the model $\beta$.  Thus the resulting network contains
$N$ nodes and $E=Eo+\Delta E$ edges, so that $\rho = E/Eo$.  The
resilience of the hybrid models was quantified by considering the
giant cluster of size $M(n)$ obtained after removing an increasing
number $n$ of nodes.  Although some analyses were performed with edge
removal (see Section Results and Discussion), the present work
concentrates on the highest degree node removal.  All simulations
adopted $No=50$ and were carried out for 100 realizations of each
configuration.

\begin{table}
\vspace{0.7cm}
\begin{tabular}{||l|c|c|c|c||} \cline{3-5} 
  \multicolumn{2}{c|}{}     &   $i=1$  &  $i=2$   &  $i=3$   \\  \hline
Random   &   $L=3$    &  $4.24 \pm 0.89$  & $5.47 \pm 0.56$  &   $7.72 \pm 0.42$    \\ \cline{3-5}
         &   $L=4$    &    $4.24 \pm 0.89$   &   $8.54 \pm 1.01$  &  $11.82 \pm 0.53$    \\ \hline
Sc.-Free &   $L=3$    &  $4.51 \pm 0.66$   &  $6.91 \pm 0.58$ &  $ 7.84 \pm 0.62 $  \\ \cline{3-5}
         &   $L=4$    &  $5.97 \pm 1.13$  &   $9.49 \pm 1.00$   & 9.50 $\pm$1.02   \\ \hline 
\end{tabular}
\caption{The values of the ratio $\rho$ for the six hybrid models considered
   in this work.~\label{tab:rho}}
\end{table}

\section{Results and Discussion}

\begin{figure*}
 \begin{center} 
   \includegraphics[scale=.7,angle=-90]{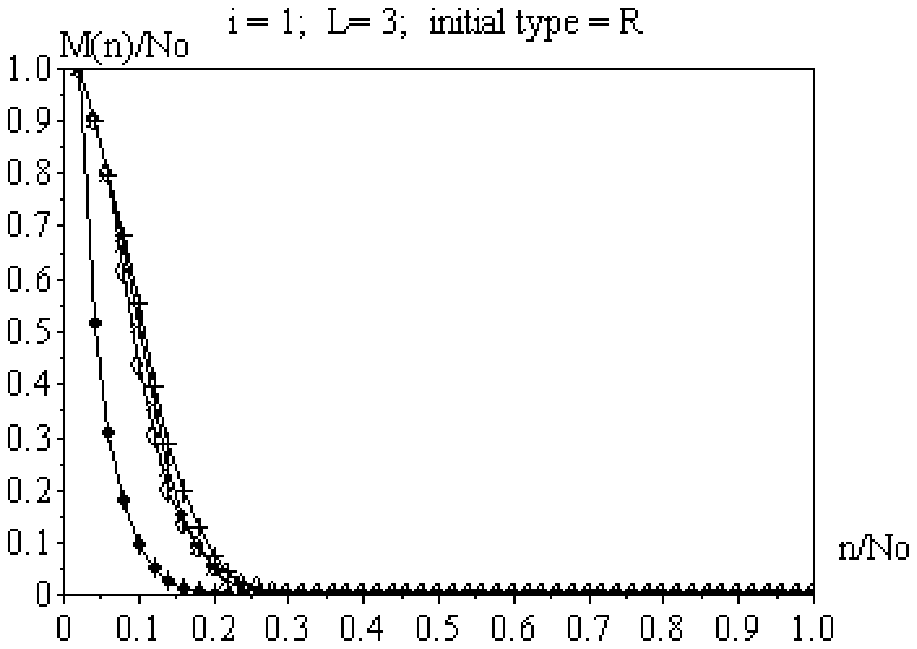} 
   \includegraphics[scale=.7,angle=-90]{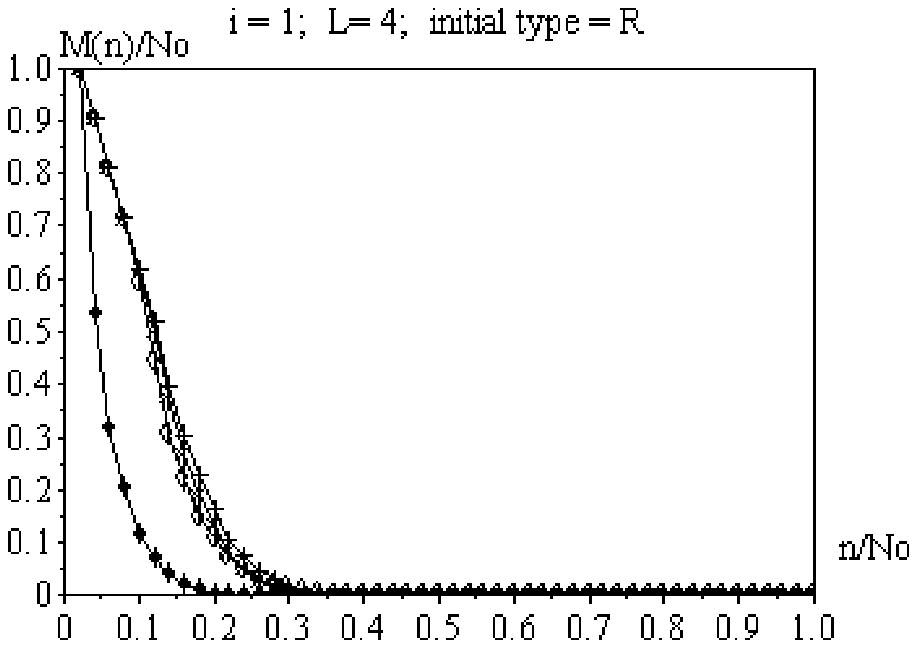} \\
   (a)  \hspace{7cm}  (b) \\

   \includegraphics[scale=.7,angle=-90]{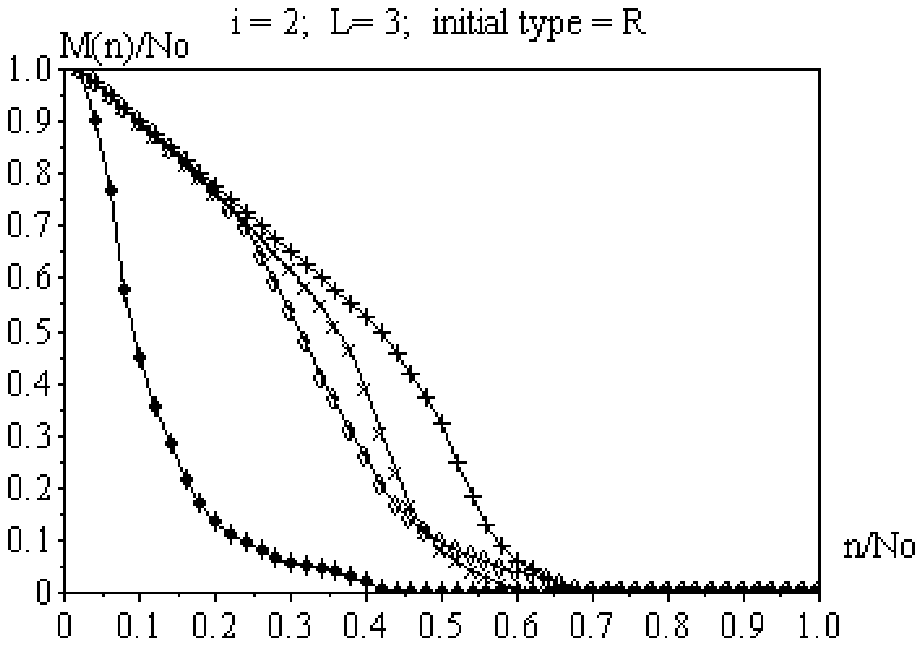} 
   \includegraphics[scale=.7,angle=-90]{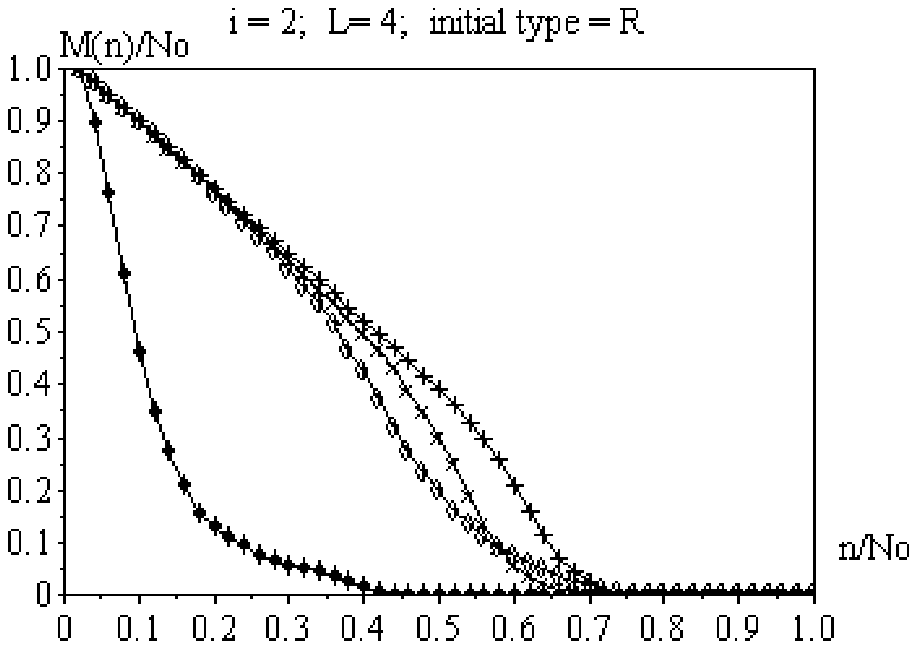} \\
   (c)  \hspace{7cm}  (d) \hspace{3cm}  \\

   \includegraphics[scale=.7,angle=-90]{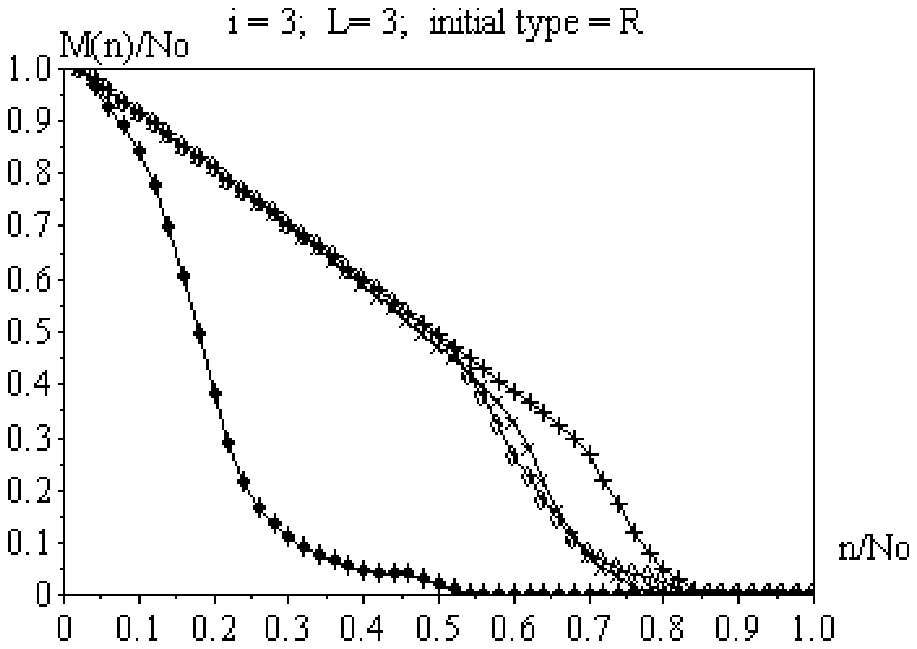} 
   \includegraphics[scale=.7,angle=-90]{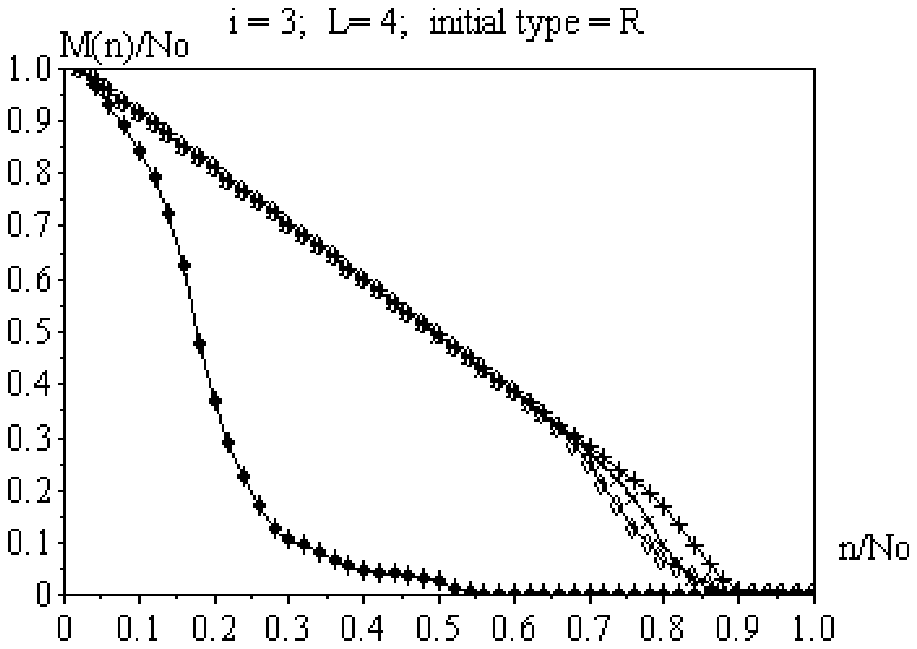} \\
   (e)  \hspace{7cm}  (f) \hspace{3cm}  \\

   \caption{The number of nodes for the cases (i)-(iii), 
         $i=1,2$ and 3 and $L=3,4$, where 
         filled $\diamond$= initial network, 
         $+$=R, $\times$=SF and $\diamond$=LE
         indicate the augmentation model.~\label{fig:res1}} \end{center}
\end{figure*}

\begin{figure*}
 \begin{center} 
   \includegraphics[scale=.7,angle=-90]{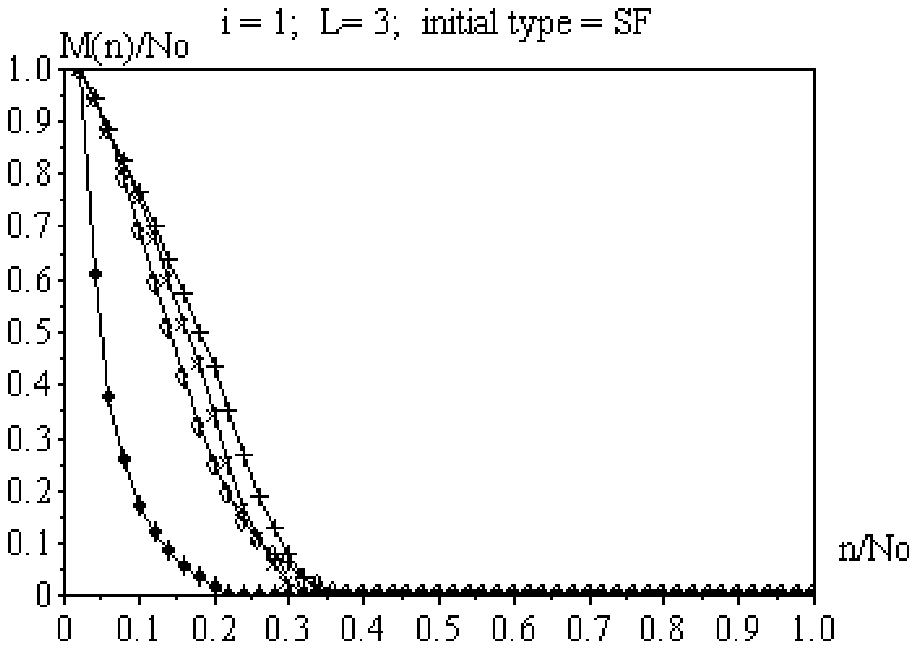} 
   \includegraphics[scale=.7,angle=-90]{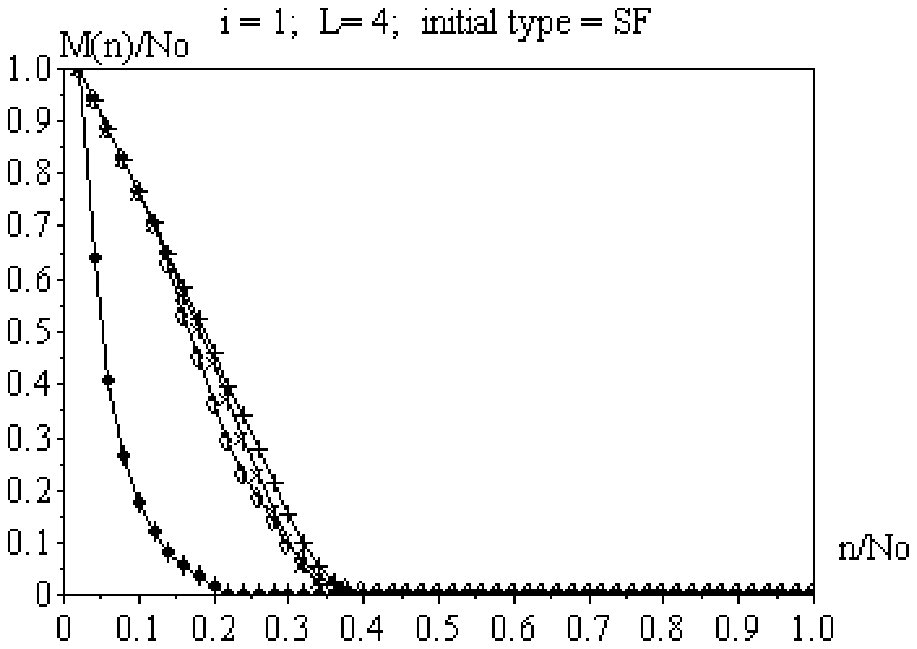} \\
   (a)  \hspace{7cm}   (b) \\

   \includegraphics[scale=.7,angle=-90]{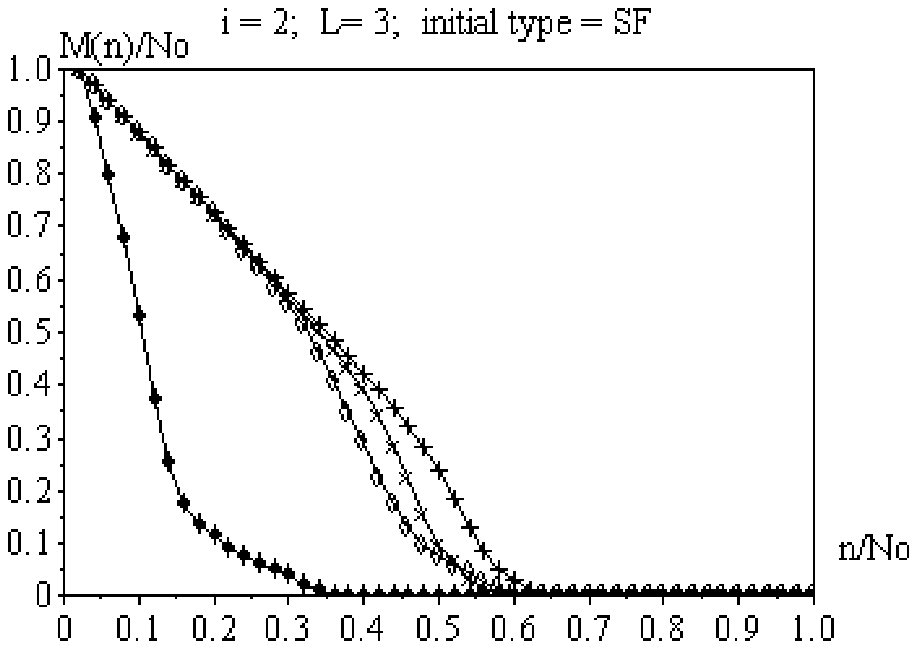} 
   \includegraphics[scale=.7,angle=-90]{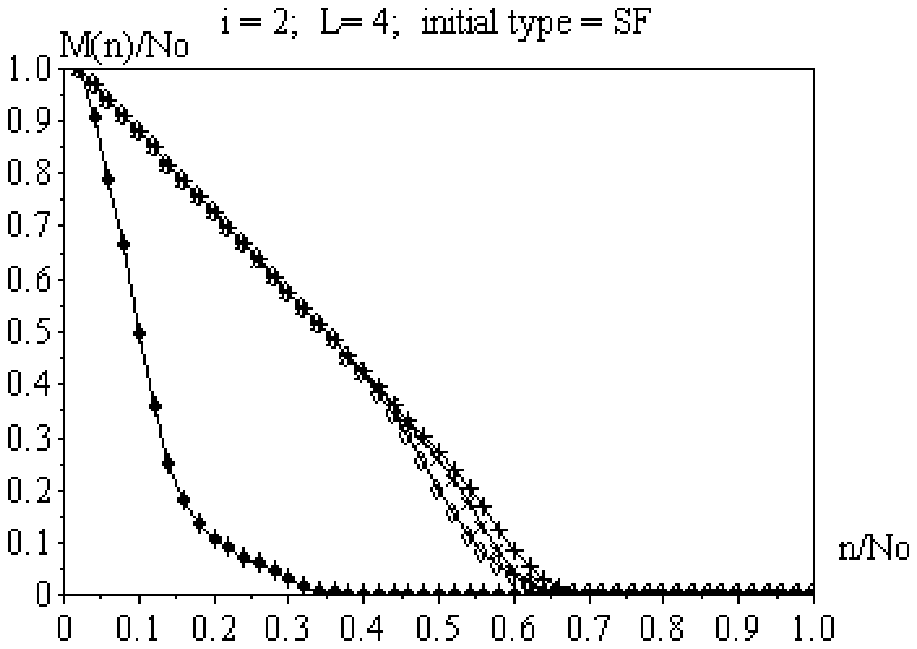} \\
   (c)  \hspace{7cm}   (d) \\

   \includegraphics[scale=.7,angle=-90]{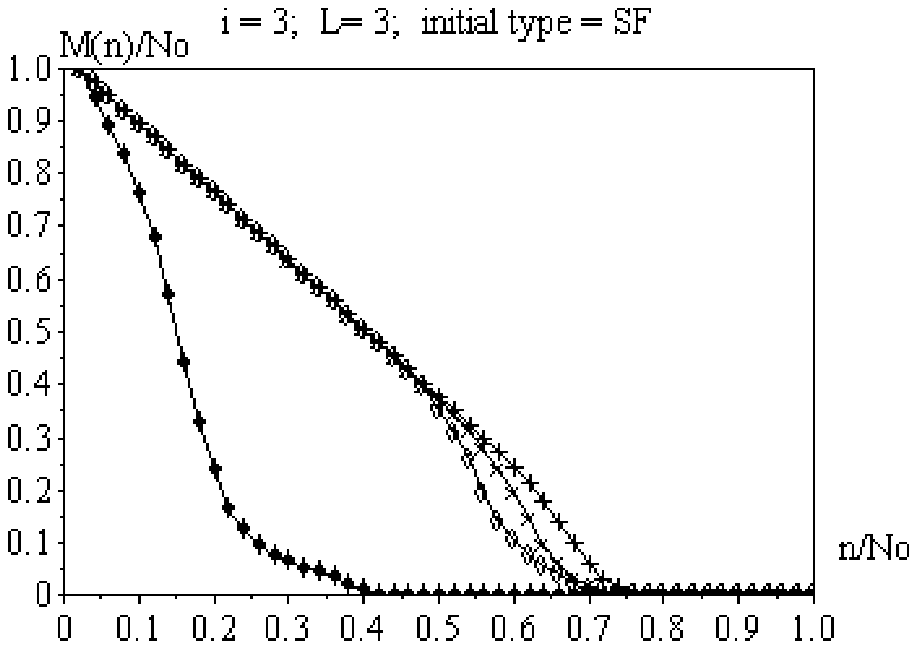} 
   \includegraphics[scale=.7,angle=-90]{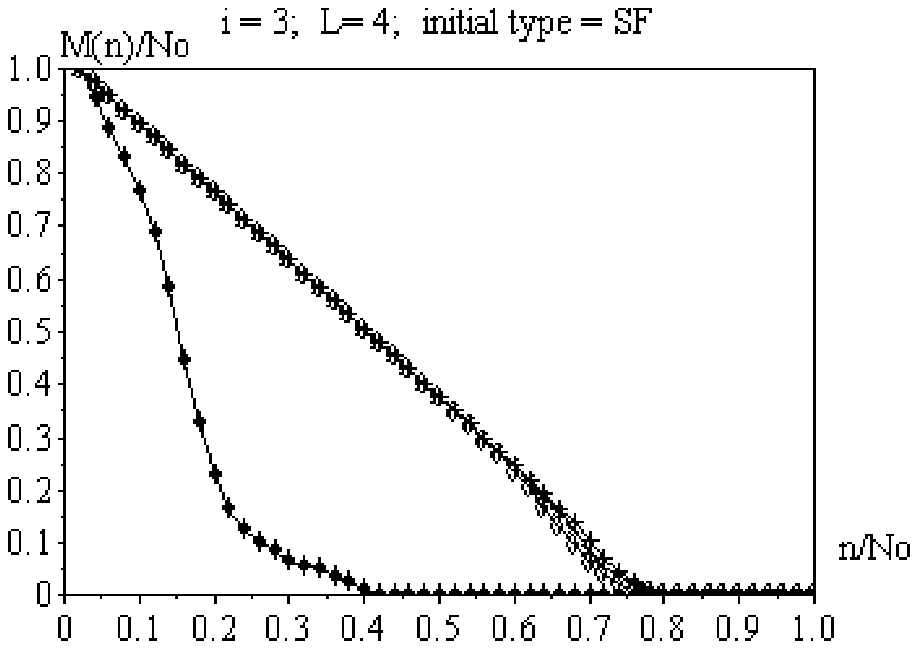} \\
   (e)  \hspace{7cm}   (f) \\

   \caption{The number of nodes for the cases (iv)-(vi), 
         $i=1,2$ and 3 and $L=3,4$, where 
         filled $\diamond$= initial network, 
         $+$=R, $\times$=SF and $\diamond$=LE indicate the
         augmentation model.~\label{fig:res2}} \end{center}
\end{figure*}

The numbers of remaining nodes in the network under attack, after $n$
removals of the nodes with maximum degree, are shown in
Figures~\ref{fig:res1} and ~\ref{fig:res2}, for random and scale-free
initial networks, respectively, and the ratios $\rho$ are shown in
Table~\ref{tab:rho}.  The growing models are identified by the curve
marks (see respective captions).  The effect of increasing values of
$i$ and, to a lesser extent of $L$, on the resilience is promptly
observed from these two figures.  In other words, the addition of
$\Delta E$ edges, quantified by the augmentation ratios in
Table~\ref{tab:rho}, contributed to substantially reinforcing the
network structure and resilience to node attacks.  The choice of
scale-free model for the initial network tended to equalize the
resilience provided by the three subsequent augmentation schemes.  The
best resilience was observed for the situation involving random
initial network adopting $i=3$ and $L=4$ (see
Figure~\ref{fig:res2}(f)), with the networking break-down occurring
only after $n/No>0.7$.  As expected \cite{Holme_etal:2002}, the random
augmentation allowed the best resilience reinforcement in \emph{all}
situations.  The \emph{L-}augmentation performed poorly, though
presenting performance superior to the scale-free model at the very
last stages of the attacks in some situations ---
i.e.~\ref{fig:res1}(b-e) and~\ref{fig:res2}(c).  Except for these
cases, the scale-free type of augmentation presented intermediate
performance.  Another interesting aspect is that the use of a
scale-free network as the initial structure implies the slope of the
decaying giant cluster size to be smaller (in absolute value) than
that of the networks augmented from random kernels. An inclination of
almost $-1$ (the highest possible, corresponding to a network where
every node is connected to every node --- i.e. \emph{complete} graph)
is obtained for large $i$ and $L$ after starting from the random
model, irrespectively to the augmentation scheme.  Investigations with
larger values of $No$ tended to produce similar results, and
simulations performed considering random edge attack seemed to
indicate little difference between the augmentation schemes.

\section{Concluding Remarks}

This paper has investigated the resilience of six hybrid network
models obtained through the process of augmenting an initial,
connected network.  This situation presents interest not only for its
theoretical implications, but also because of practical concerns while
trying to enhance the design of specific network systems in order to
suit fault-tolerance specifications.  The six network models included
the random and scale-free traditional networks augmented by random,
scale-free and \emph{L-}augmentations (for $L=3$ and 4).  The obtained
results revealed some interesting aspects.  First, the augmentation of
the initial giant cluster was observed to substantially enhance the
resilience of the final network, at the expense of a larger number of
edges.  Second, the random model confirmed its superiority regarding
the highest degree nodes attack, with the scale-free networks coming
second, except at the very last stages of the attack sequence, where
the \emph{L-}augmentations tended to provide behaviour similar to that
of the random networks.  Another interesting result is the fact that
the initial model (type and growth parameters) determines to a great
extent the possibilities for reinforcing the initial network.  The
obtained results corroborated the resilience superiority of fully
random connection when used both as the initial and as the
augmentation model.  

The problem of reinforcement can be understood as a specific situation
of a broader class of problems where one wants to \emph{redesign} or
\emph{adapt} a given network in order to obtain specific topological
or functional properties.  Such a situation could arise in several
contexts, for instance in internetworking, electronic circuits
(analogic or digital), and also biology.  Regarding this latter
situation, a particularly interesting case is the exposal of existing
biological networks --- including metabolic, protein, food chain and
ecologic -- to abrupt environmental variations of the geographical,
environmental and metereologic conditions that permeate the
evolutionary process.

Future works may consider the evaluation of the performance of other
hybrid systems, such as those obtained by union of two distinct models
(e.g. \cite{Costa_topo:2003}), progressive modification of the growing
scheme (e.g. \cite{Costa_etal:2003}), or even successive alternations
of the augmentation schemes.  Another interesting further work is to
devise modifications of the \emph{L-}augmentation scheme where
augmentations are not applied indiscriminately over all nodes, but at
random or selectively to specially critical nodes (e.g. those with low
degree or betweeness centrality).  Actually, such a line of reasoning
inevitably leads to the following question:

\begin{trivlist}

  \item $\ast$ \emph{Given an existing network, how to identify the
optimal augmentation scheme, i.e. that leading to the best overall
resilience at the expense of the smallest number of additional edges?}

\end{trivlist}

Although this type of problem has been well-developed in the context
of traditional graphs (e.g.\cite{Frank:1994}), it would be interesting
to revisit it by considering the new concepts and results from complex
network research.  Another related question is:

\begin{trivlist}

  \item $\ast$ \emph{Given an augmented network, how to identify the
initial and/or expanding models?}

\end{trivlist}

For instance, the concept of \emph{L-}conditional expansion
\cite{Costa_expand:2004} can be used to identify the regular
connections implied by \emph{L-}augmentations.  In addition, it is
worth observing that, in addition to enhancing the connectivity of the
initial network, \emph{L-}augmentation are also likely to promote
higher regularity of node degree at the scale defined by $L$.
Possible means to identify augmentation schemes leading to high (or
maximum) resilience is to use simulated annealing or the genetic
algorithm.

\begin{acknowledgments}

The author is grateful to FAPESP (proc. 99/12765-2) and CNPq
(proc. 301422/92-3) for financial support.

\end{acknowledgments}

 
\bibliography{resil}

\begin{thebibliography}{17}
\expandafter\ifx\csname natexlab\endcsname\relax\def\natexlab#1{#1}\fi
\expandafter\ifx\csname bibnamefont\endcsname\relax
  \def\bibnamefont#1{#1}\fi
\expandafter\ifx\csname bibfnamefont\endcsname\relax
  \def\bibfnamefont#1{#1}\fi
\expandafter\ifx\csname citenamefont\endcsname\relax
  \def\citenamefont#1{#1}\fi
\expandafter\ifx\csname url\endcsname\relax
  \def\url#1{\texttt{#1}}\fi
\expandafter\ifx\csname urlprefix\endcsname\relax\def\urlprefix{URL }\fi
\providecommand{\bibinfo}[2]{#2}
\providecommand{\eprint}[2][]{\url{#2}}

\bibitem[{\citenamefont{Albert and Barab\'asi}(2002)}]{Albert_Barab:2002}
\bibinfo{author}{\bibfnamefont{R.}~\bibnamefont{Albert}} \bibnamefont{and}
  \bibinfo{author}{\bibfnamefont{A.~L.} \bibnamefont{Barab\'asi}},
  \bibinfo{journal}{Rev. Mod. Phys.} \textbf{\bibinfo{volume}{74}},
  \bibinfo{pages}{47} (\bibinfo{year}{2002}).

\bibitem[{\citenamefont{Newman}(2003)}]{Newman:2003}
\bibinfo{author}{\bibfnamefont{M.~E.~J.} \bibnamefont{Newman}},
  \bibinfo{journal}{SIAM Review} \textbf{\bibinfo{volume}{45}},
  \bibinfo{pages}{167} (\bibinfo{year}{2003}),
  \bibinfo{note}{cond-mat/0303516}.

\bibitem[{\citenamefont{Dorogovtsev and Mendes}(2002)}]{Dorog_Mendes:2002}
\bibinfo{author}{\bibfnamefont{S.~N.} \bibnamefont{Dorogovtsev}}
  \bibnamefont{and} \bibinfo{author}{\bibfnamefont{J.~F.~F.}
  \bibnamefont{Mendes}}, \bibinfo{journal}{Advances in Physics}
  \textbf{\bibinfo{volume}{51}}, \bibinfo{pages}{1079} (\bibinfo{year}{2002}),
  \bibinfo{note}{cond-mat/0106144}.

\bibitem[{\citenamefont{Erd\H{o}s and R\'enyi}(1959)}]{Erdos:1959}
\bibinfo{author}{\bibfnamefont{P.}~\bibnamefont{Erd\H{o}s}} \bibnamefont{and}
  \bibinfo{author}{\bibfnamefont{A.}~\bibnamefont{R\'enyi}},
  \bibinfo{journal}{Pub. Math.} \textbf{\bibinfo{volume}{6}},
  \bibinfo{pages}{290} (\bibinfo{year}{1959}).

\bibitem[{\citenamefont{Watts and Strogatz}(1998)}]{Watts_Strogatz:1998}
\bibinfo{author}{\bibfnamefont{D.~J.} \bibnamefont{Watts}} \bibnamefont{and}
  \bibinfo{author}{\bibfnamefont{S.~H.} \bibnamefont{Strogatz}},
  \bibinfo{journal}{Nature} \textbf{\bibinfo{volume}{393}},
  \bibinfo{pages}{440} (\bibinfo{year}{1998}).

\bibitem[{\citenamefont{Pastor-Satorras and Vespignani}(2001)}]{Pastor:2001}
\bibinfo{author}{\bibfnamefont{R.}~\bibnamefont{Pastor-Satorras}}
  \bibnamefont{and}
  \bibinfo{author}{\bibfnamefont{E.}~\bibnamefont{Vespignani}},
  \bibinfo{journal}{Phys. Rev. Lett.} \textbf{\bibinfo{volume}{86}},
  \bibinfo{pages}{3200} (\bibinfo{year}{2001}),
  \bibinfo{note}{cond-mat/0010317}.

\bibitem[{\citenamefont{da~F.~Costa}(2004)}]{Costa_expand:2004}
\bibinfo{author}{\bibfnamefont{L.}~\bibnamefont{da~F.~Costa}}
  (\bibinfo{year}{2004}), \bibinfo{note}{cond-mat/0312712}.

\bibitem[{\citenamefont{Albert et~al.}(2000)\citenamefont{Albert, Jeong, and
  Barab\'asi}}]{Albert_attack:2000}
\bibinfo{author}{\bibfnamefont{R.}~\bibnamefont{Albert}},
  \bibinfo{author}{\bibfnamefont{H.}~\bibnamefont{Jeong}}, \bibnamefont{and}
  \bibinfo{author}{\bibfnamefont{A.~L.} \bibnamefont{Barab\'asi}},
  \bibinfo{journal}{Nature} \textbf{\bibinfo{volume}{406}},
  \bibinfo{pages}{378} (\bibinfo{year}{2000}),
  \bibinfo{note}{cond-mat/0008064}.

\bibitem[{\citenamefont{Callaway et~al.}(2000)\citenamefont{Callaway, Newman,
  Strogatz, and Watts}}]{Callaway_etal:2000}
\bibinfo{author}{\bibfnamefont{D.~S.} \bibnamefont{Callaway}},
  \bibinfo{author}{\bibfnamefont{M.~E.~J.} \bibnamefont{Newman}},
  \bibinfo{author}{\bibfnamefont{S.~H.} \bibnamefont{Strogatz}},
  \bibnamefont{and} \bibinfo{author}{\bibfnamefont{D.~J.} \bibnamefont{Watts}},
  \bibinfo{journal}{Phys. Rev. Lett.} \textbf{\bibinfo{volume}{85}},
  \bibinfo{pages}{5468} (\bibinfo{year}{2000}),
  \bibinfo{note}{cond-mat/0007300}.

\bibitem[{\citenamefont{Holme et~al.}(2002)\citenamefont{Holme, Kim, Yoon, and
  Han}}]{Holme_etal:2002}
\bibinfo{author}{\bibfnamefont{P.}~\bibnamefont{Holme}},
  \bibinfo{author}{\bibfnamefont{B.~J.} \bibnamefont{Kim}},
  \bibinfo{author}{\bibfnamefont{C.~N.} \bibnamefont{Yoon}}, \bibnamefont{and}
  \bibinfo{author}{\bibfnamefont{S.~K.} \bibnamefont{Han}},
  \bibinfo{journal}{Phys. Rev. E} \textbf{\bibinfo{volume}{65}},
  \bibinfo{pages}{056109} (\bibinfo{year}{2002}),
  \bibinfo{note}{cond-mat/0202410}.

\bibitem[{\citenamefont{Cohen et~al.}(2000)\citenamefont{Cohen, Erez, ben
  Avraham, and Havlin}}]{Cohen_etal:2000}
\bibinfo{author}{\bibfnamefont{R.}~\bibnamefont{Cohen}},
  \bibinfo{author}{\bibfnamefont{K.}~\bibnamefont{Erez}},
  \bibinfo{author}{\bibfnamefont{D.}~\bibnamefont{ben Avraham}},
  \bibnamefont{and} \bibinfo{author}{\bibfnamefont{S.}~\bibnamefont{Havlin}},
  \bibinfo{journal}{Phys. Rev. Lett.} \textbf{\bibinfo{volume}{85}},
  \bibinfo{pages}{4626} (\bibinfo{year}{2000}),
  \bibinfo{note}{cond-mat/0007048}.

\bibitem[{\citenamefont{Newman et~al.}(2002)\citenamefont{Newman, Forrest, and
  Balthrop}}]{Newman:2002}
\bibinfo{author}{\bibfnamefont{M.~E.~J.} \bibnamefont{Newman}},
  \bibinfo{author}{\bibfnamefont{S.}~\bibnamefont{Forrest}}, \bibnamefont{and}
  \bibinfo{author}{\bibfnamefont{J.}~\bibnamefont{Balthrop}},
  \bibinfo{journal}{Phys. Rev. E} \textbf{\bibinfo{volume}{66}},
  \bibinfo{pages}{035102} (\bibinfo{year}{2002}).

\bibitem[{\citenamefont{Jeong et~al.}(2000)\citenamefont{Jeong, Tombor, Albert,
  Oltvai, and Barab\'asi}}]{Jeong_etal:2000}
\bibinfo{author}{\bibfnamefont{H.}~\bibnamefont{Jeong}},
  \bibinfo{author}{\bibfnamefont{B.}~\bibnamefont{Tombor}},
  \bibinfo{author}{\bibfnamefont{R.}~\bibnamefont{Albert}},
  \bibinfo{author}{\bibfnamefont{Z.~N.} \bibnamefont{Oltvai}},
  \bibnamefont{and} \bibinfo{author}{\bibfnamefont{A.~L.}
  \bibnamefont{Barab\'asi}}, \bibinfo{journal}{Nature}
  \textbf{\bibinfo{volume}{407}}, \bibinfo{pages}{651} (\bibinfo{year}{2000}),
  \bibinfo{note}{cond-mat/0010278}.

\bibitem[{\citenamefont{Dunne et~al.}(2002)\citenamefont{Dunne, Williams, and
  Martinez}}]{Dunne:2002}
\bibinfo{author}{\bibfnamefont{J.~A.} \bibnamefont{Dunne}},
  \bibinfo{author}{\bibfnamefont{R.~J.} \bibnamefont{Williams}},
  \bibnamefont{and} \bibinfo{author}{\bibfnamefont{N.~D.}
  \bibnamefont{Martinez}}, \bibinfo{journal}{Proc. Natl. Acad. Sci. USA}
  \textbf{\bibinfo{volume}{99}}, \bibinfo{pages}{12917} (\bibinfo{year}{2002}).

\bibitem[{\citenamefont{Frank}(1994)}]{Frank:1994}
\bibinfo{author}{\bibfnamefont{A.}~\bibnamefont{Frank}},
  \emph{\bibinfo{title}{Mathematical Programming: State of the Art}}
  (\bibinfo{publisher}{The University of Michigan, J. R. Birge and K. G.
  Murty}, \bibinfo{year}{1994}), chap. \bibinfo{chapter}{Connectivity
  augmentation}, pp. \bibinfo{pages}{34--63}.

\bibitem[{\citenamefont{da~F.~Costa}(2003)}]{Costa_topo:2003}
\bibinfo{author}{\bibfnamefont{L.}~\bibnamefont{da~F.~Costa}}
  (\bibinfo{year}{2003}), \bibinfo{note}{cond-mat/0306530}.

\bibitem[{\citenamefont{da~F.~Costa et~al.}(2003)\citenamefont{da~F.~Costa,
  Travieso, and Ruggiero}}]{Costa_etal:2003}
\bibinfo{author}{\bibfnamefont{L.}~\bibnamefont{da~F.~Costa}},
  \bibinfo{author}{\bibfnamefont{G.}~\bibnamefont{Travieso}}, \bibnamefont{and}
  \bibinfo{author}{\bibfnamefont{C.~A.} \bibnamefont{Ruggiero}}
  (\bibinfo{year}{2003}), \bibinfo{note}{cond-mat/0312603}.

\end{thebibliography}

\end{document}